# Who's Who in the Information Technology Research in the Philippines

## A Social Network Analysis

Rex P. Bringula
College of Computer Studies and
Systems, University of the East
Philippines
rex.bringula@ue.edu.ph

Ma. Carmela Racelis
Adamson University
Philippines
ma.carmela.racelis@adamson.ed
u.ph

Rey C. Rodrigueza
Information and
Communications Technology
Department, Sorsogon State
University-Bulan Campus,
Philippines
rcrodrigueza@gmail.com

## ABSTRACT

This study reported the conference papers presented conducted by the two computing societies in the Philippines. Toward this goal, all published conference proceedings from the National Conference of IT Education and Philippine Computing Society Conference were gathered and analyzed using social network analysis. The findings of the study disclosed that there are 733 papers presented in the conference for the span of 18 years. On the average, both conferences had 27 papers presented annually. Private higher education institutions dominated the list of research productive schools where De La Salle University tops the list. A researcher in the University of the Philippines-Diliman is the most prolific researcher with 39 publications and "algorithm" was the most researched topic. Researchers tend to work in small team consisting of 2 to 3 members. Implications and limitations of the study are also presented.

## CCS CONCEPTS

• Human Factors

## 1 Introduction

Two professional organizations support the advancement of computing education (i.e., computer science, information technology, information systems, multimedia computing, and game development) in the Philippines. The Philippine Society of Information Technology Educators (PSITE), which was instituted in 1998, aims to "promote quality IT education in the Philippines through workshops, seminars, industry-academe linkages, and programs designed to benefits its members, and the academic sector in general" [9]. Meanwhile, the Computing Society of the Philippines (CSP) is a professional organization of computing researchers and educators in the Philippines that aims to promote research and development in computing science. CSP was founded in 2005 [3].

To meet this common objective, both organizations have been conducting annual national conferences for more than 10 years. With the exception in 2002, the CSP has been conducting annual conference called Philippine Computing Science Congress (PCSC) since 2000. There are 18 conference proceedings since 2000. Meanwhile, PSITE holds the National Convention on IT Education (NCITE) since 2003 [9]. To date, no study has been conducted to understand the contributions of the two national conferences in computing research in the Philippines.

Toward this goal, this study reported the state of Information Technology (IT) research in the Philippines. Conference papers published by the two societies were gathered and analyzed through social network analysis. Specifically, the study sought answers to the following questions. 1) How can we describe the state of IT research in terms of total publications, publications per institution, publications per researcher, authorship, and research areas investigated? 2) How can we describe the research productivity of most published researchers through social network analysis?

## 2 Literature Review

### 2.1 Research Productivity

Research productivity is defined as the research outputs in forms of journal articles, patents, conference proceedings, research fundings, and other scholarly works [8]. Different studies were conducted that described research productivity. In Israel, Naser-Abu Alhija and Majdob [7] found that academic degree, rank, administrative position, desire to develop new knowledge and learn from research findings, and perceived insufficient research competence and self-confidence predicted teachers' research productivity. The authors suggested that these variables be considered when recruiting teachers, assigning administrative responsibilities, and instituting professional development programs.

Obuku et al. [8] conducted a study a systematic review on research productivity in the field of health policy of post-graduate students in low- and middle-income countries (e.g., Brazil, Cameroon, Egypt, India, Iran, Peru, Togo, Turkey, Uganda, Zambia). The primary purpose of the study is twofold. First, they determined the post-graduate students' research productivity or the application of post-graduate students' research. Second, they aimed to find studies that assessed the determinants that contributed to the productivity and use of the post-graduate students' research. Out of the 5,080 candidate published articles retrieved from PubMed and ERIC, only 44 articles were analyzed. Articles published earlier than 1990 were excluded. They found that there is a low research productivity of post-graduate students in low- and middle-income countries. They also discovered that there is no study that assessed the strategies that increased productivity and use of post-graduate students' research.

Allen et al. [1] identified the factors that explained the research productivity of African-American social work faculty. The researchers conducted qualitative interviews with 10 top-ranked African-American faculty. They found that mentorship, collaboration, time, and strategic planning were the factors that increased research productivity. Of these factors, mentorship was the most prominent factor. The researchers suggested that faculty members aiming to increase research productivity may utilize the findings of the study.

Kuzhabekova and Ruby [4] reported the strategy of the government of Kazakhstan in increasing faculty research productivity. The government-initiated policy required university faculty members to publish in journals with impact factors as a requirement for promotion. They investigated the impact of this policy in six universities in the said country. They found this strategy is effective because support structures are in place and universities are able to control the promotion process.

Lou, Wang and Yang [6] compared the research productivity and impact of Chinese scholars in China and overseas. Toward this goal, they investigated the research publications of 1,190 Chinese scholars in China and of 1,983 Chinese scholars overseas. A total of 6,306 papers published over the 10 years in 6 journals were analyzed. Average number of authors, publications, citations, and usage counts were the indicators of research productivity and impact. Statistical analyses (e.g., standard deviation and covariance analysis) disclosed that Chinese scholars in their homeland were more productive and have impactful research than those of their counterparts conducting research abroad.

In a similar study, Vuong et al. [12] analyzed the effects of work environment and collaboration on research productivity of social sciences researchers in Vietnam. Published articles indexed in Scopus ranging from 2008 to 2017 were collected. Ordinary least squares method showed that university-affiliated authors in Vietnam had higher research productivity than their institution-affiliated peers. International collaboration could increase research productivity but it had insignificant effect among high-performing researchers.

Lastly, Lase and Hartijasti [5] analyzed the effects of individual (e.g., socialization, motivation, work habits, etc.) and leadership characteristics leadership characteristics (e.g., scholarship, research orientation, capability to fulfill all critical leadership roles, and active leadership participation) on research productivity on lecturers' research productivity in one university in Indonesia. They also included institutional characteristics (e.g., resources, rewards, sufficient work time, etc.) in the analysis as a mediating factor. It is revealed that the relationship between individual characteristics and research productivity is partially mediated by institutional characteristics. On the other hand, full mediation was found between leadership characteristics and research productivity.

### 2.2 Social Network Analysis

Social network analysis (SNA) is a method of mapping and measuring relationships among people, groups, and any source of information or knowledge that is connected to one another [2]. SNA is applied in various fields of study. For example, Rodrigueza and Estuar [10] used SNA and agent-based modeling to analyze the behavior of people during disasters. They found that there are agents in the network that played a significant role during disaster risk reduction and management.

SNA was also utilized in analyzing the interaction of online forum users [11]. Through this method, they were able to identify students that actively and not actively participating in the educational online forum. Another study utilized SNA to identify students and teachers that need academic assistance. This paper used SNA because traditional educational evaluations did not provide a clear understanding of academic exchange between and among students and teachers. While it is expected that teachers are the primary source of academic assistance for students, it is worth noting that more-knowledgeable individuals do not necessarily provide more assistance to those who need it.

## 2.3 Synthesis

It can be observed that the above prior studies are focused on finding the factors that influence research productivity and identifying the indicators that are used to measure research productivity. The studies of Naser-Abu Alhija and Majdob [7], Allen et al. [1], Vuong et al. [12], and Lase and Hartijasti [5] can be classified in the first goal. Their collective findings suggest that individual-, institution-, and government-related factors influence research productivity. On one hand, the study of Obuku et al. [8] and Lou, Wang and Yang [6] attempted to measure research productivity. Number of citations, number of published articles, and usage counts were some of the indicators used in measuring research productivity.

This paper intends to contribute to the existing literature by describing research productivity of Filipino researchers. However, it does not intend to find the factors that contribute to research productivity. Instead, it describes research productivity of Filipino researchers in terms of total publications, publications per institution, publications per researcher, authorship, and research areas investigated. Moreover, SNA is utilized which could provide further insights on the research productivity of Filipino researchers.

## 3 Methodology

### 3.1 Sources of Data, Data Collection Procedure, Sample Size, and Sampling Design

All conference proceedings of the NCITE and PCSC were collected in the official conference websites of the societies [https://sites.google.com/a/dcs.upd.edu.ph/csp-proceedings/]. The CSP had 18 online conference proceedings since 2000. NCITE had nine online conference proceedings since 2009 [https://sites.google.com/site/phncite/2009]. The name of the authors, title of the paper, year of publication, affiliation, and concepts were manually extracted from the proceedings. The two societies manage two journals (Philippine Information Technology Journal and Philippine Computing Journal). Publications in these journals are excluded because there are delays in publications. Conference proceedings are regularly published annually, which could reflect the actual state of the IT research in the Philippines. All attributes were manually encoded in a spreadsheet. Seven hundred and ten papers were collected and they were all considered in the analysis.

### 3.2 Data Cleaning, Data Preprocessing, and Data Analysis

The collected data was subjected to data cleaning. All special characters were removed (e.g., ":", "@"). Afterwards, the dataset was subjected to data preprocessing. Stopwords comprising of English standard terms (e.g., "the", "a", "an",

"using", etc.), keyword terms (e.g., "system", "analysis"), and noise words (e.g., alphanumeric, whitespace, punctuations, and unrecognizable characters) were removed in the dataset. All words in the dataset were transformed into lowercase and tokenized that comprised the corpus. The words were subjected to stemming. Then, bigrams were generated to have interpretable words. The corpus contained 2,281 words. RapidMiner was utilized in processing the dataset.

There are two methods of data analysis in this study. The first method analyzed the corpus through descriptive statistics (e.g., frequency counts and percentages). This method aims to answer the first research question. The second analysis involved social network analysis (SNA). ORA-Lite was utilized in conducting the SNA. Total-degree centrality measure was used to determine the consistency of the publication of the most published authors and the number of collaborators (e.g., single, 2 authors, 3 authors, etc).

## 4. Results

**Research Question 1**

There are 733 papers presented in NCITE and PCSC. One hundred and forty-three (143) schools participated in these conferences. For the span of 9 years, NCITE had 239 conference papers. Meanwhile, 494 papers were presented in PCSC for the span of 18 years. This means that, on average, both conferences had 27 papers presented every year.

De La Salle University (n = 140) had the highest number of papers in the conference followed by the University of the Philippines-Diliman (UPD, n = 128). Ateneo de Manila University (ADMU, n = 103) had the third highest number of contributing authors. The University of the Philippines-Los Banos (UPLB) ranked fourth with 93 authors and the Technological Institute of the Philippines (TIP) ranked 5th with 25 authors. The other half of the list is consisted of Ateneo de Naga (6th rank, n = 20), Mindanao State University-Iligan (7th rank, n = 18), Siliman University (8th rank, n = 11), Cebu Institute of Technology (CIT) (9th rank, n = 10), and Mapua Institute of Technology (Mapua) and Ateneo de Davao University (10th rank, n = 8) (Figure 1).

It is worth noting that the first-five institutions were also had the most number of authors in the conference (Figure 2). MSU-Iligan, Ateneo de Naga, Mapua Institute of Technology, and Siliman University were also active institutions in terms of number of authors. It can be observed in Figure 2 that CIT and Ateneo de Davao are not included in Figure 2. Furthermore, University of San Carlos (USC) is in Figure 1 but not in Figure 2. This is because the first-two HEIs had only 14 and 19 authors, respectively. On the other hand, USC had only 5 papers but has 20 authors.

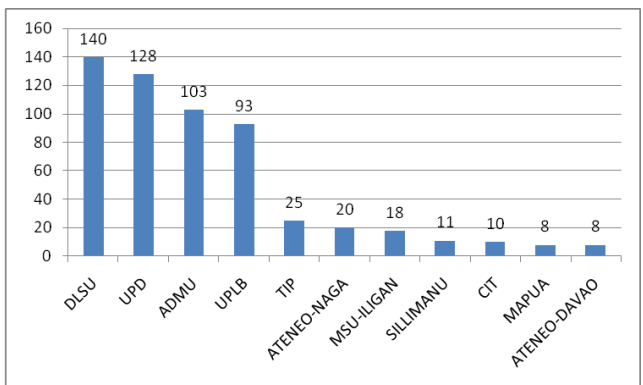

**Figure 1. Top 10 HEIs with Most Number of Papers in NCITE and PCSC**

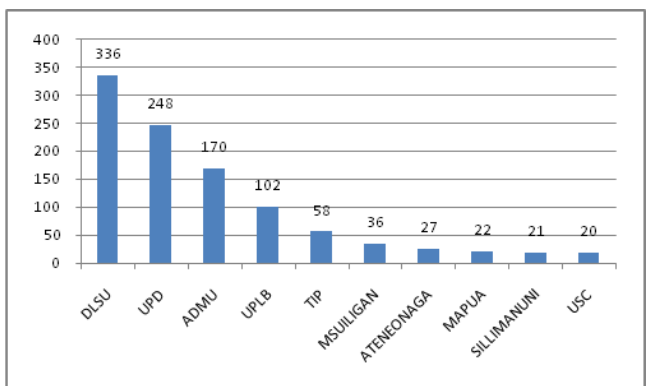

**Figure 2. Top 10 HEIs with the Most Number of Authors in NCITE and PCSC**

Henry Adorna of UPD had the most number of papers (n = 39) presented in the conference (Figure 3). He had at least presented two papers per year. Jaderick Pabico and Vladimir Mariano, both from UPLB, had the second and third most paper presentations, respectively. Prospero Naval Jr. of ADMU is on the fourth spot. Allan Sioson of Ateneo de Naga and Proceso Fernandez of ADMU shared the fifth spot as both of them had 17 papers. Merlin Suarez (DLSU, n = 15), Rafael Saldana (ADMU, n = 14), Ma. Mercedes Rodrigo (ADMU, n = 13), Jasmine Malinao (UPD, n = 12), Richelle Juayong (UPD, n = 12), Raymund Sison (DLSU, n = 12), Rachel Roxas (DLSU,NU, n = 12), and Elmer Maravillas (CIT, n = 11) completed the list of most published researchers. There are 13 authors identified as most published researchers. Their combined publications (244) are more than one-third of the total publications.

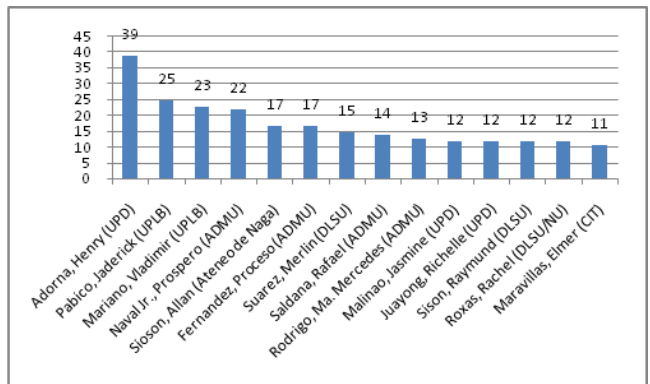

**Figure 3. Most Published Researchers**

Based on the author-supplied ACM Computing Classification System concepts, the most investigated topic is about algorithms (n = 83). There is a wide range of topics investigated in the conference. The research topics investigated are in the field of design, systems, management, measurement, computing, experimentation, data, information, data, factors, computer, human factors, and mobile.

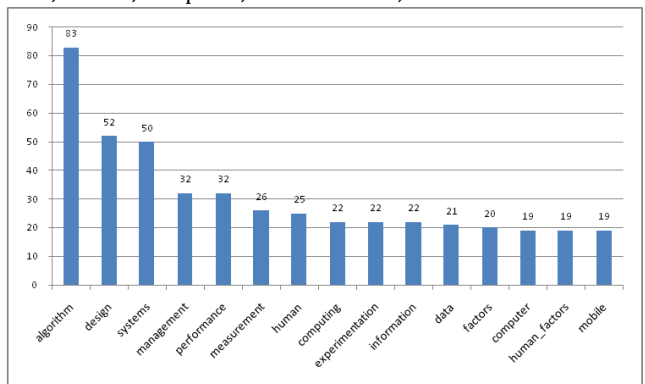

**Figure 4. Most Researched Topic**

More than one-third of the papers had two authors (n = 220, 32%). Twenty percent (n = 137) of the papers had at least 5 authors. There is equal number of authors for single (n = 105, 15%) and four (n = 105, 15%) authorships. Eighteen percent (n = 126) of the papers had 3 authors. It is clear that researchers prefer to collaborate with their peers.

Table 1. Number of Authors per Paper

| Authorship | Frequency | % |
|---|---|---|
| 1 author | 105 | 15 |
| 2 authors | 220 | 32 |
| 3 authors | 126 | 18 |
| 4 authors | 105 | 15 |
| at least 5 authors | 137 | 20 |
| Total | 693 | 100% |

**Research Question 2**

Using total-degree centrality measure, Figure 5 shows that the largest nodes in red color were assigned to the researchers Henry Adorna and Allan Sioson. This means that they are the most consistent authors in contributing to the conferences. Furthermore, it is revealed that most of the authors (11 out of 13) were all active in submitting papers during 2009 and 2011 (nodes in green color).

The author-by-number-of-collaborators analysis disclosed that the 13 authors tend to form a team of researchers with 2 or 3 members. The same centrality measure was utilized in finding the largest node. It can be infer that the researchers prefer to work in a small group.

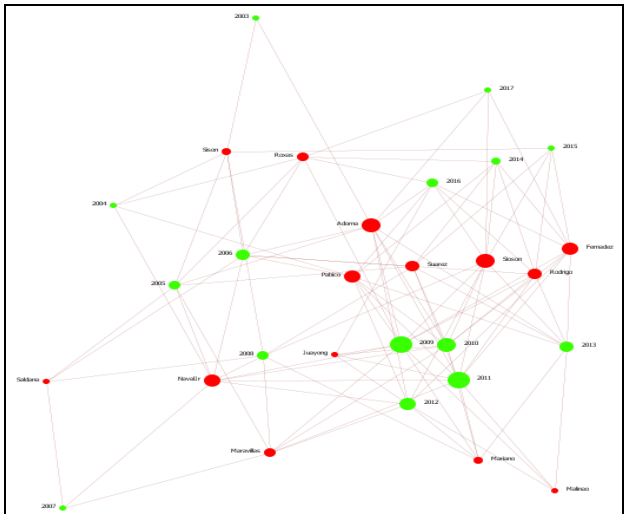

**Figure 5. Author-by-Year Analysis**

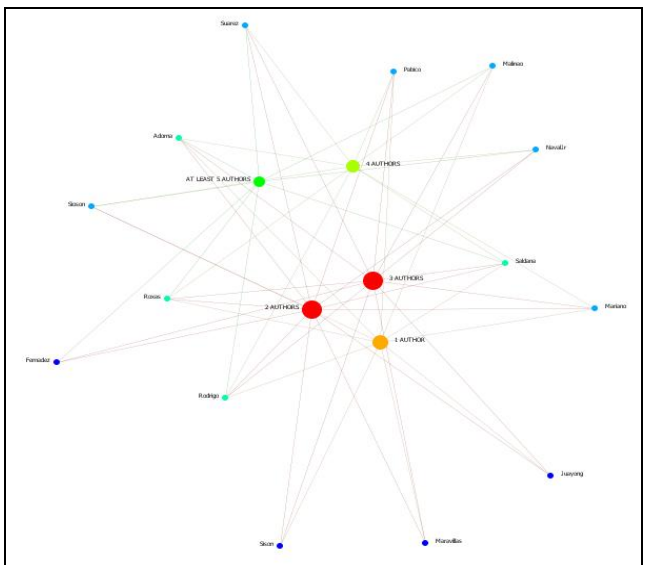

**Figure 6. Author-by-Number of Collaborators Analysis**

# 5. Discussion

This study intends to describe the state of IT research in the Philippines. Conference proceedings of two academic professional organizations in the Philippines are gathered and analyzed. The total number of papers and the average number of papers accepted in both conferences implies that both organizations are able attract submissions from the Philippine higher education institutions (PHEIs). This can be explained by the fact that these are the only two professional organizations in the Philippines that conduct such conferences.

The findings signify that the PHEIs are participating actively in conducting IT research, which in turn, indicates that PHEIs are responding on the need to develop a culture of IT research in the Philippines. As shown in the findings, DLSU is the leading institution that responds to this endeavor. It is the most active institution both in terms of quantity of papers and authors. DLSU is a private HEI. ADMU is another private HEI that is in the forefront of IT research. DLSU, UPD, ADMU, and UPLB had at least 90 papers and had at least 100 authors.

Three state-run universities are included in the list – UPD, UPLB and MSU-Iligan. UPD and UPLB had second and fourth in terms of number of papers and number of authors. Their combined contributions could overtake the contributions of DLSU. Nonetheless, it is apparent that the state-owned university is also the leading PHEIs in IT research.

TIP, Silliman University, Ateneo de Naga, CIT, Ateneo de Davao, and Mapua are private HEIs with most contributions in the conference. It can be observed that the list is dominated by private HEIs. The findings also provided an insight that there is still a large disparity of contributions from the HEIs. For instance, UPLB, which is on the fourth rank, has 68 more papers than that of TIP.

Henry Adorna of UPD is the most prolific researcher. He has contributed 2 conference papers per year. He and Allan Sioson are contributing to the conference consistently. It can be observed that the most published researchers are affiliated with UPD, UPLB, ADMU, Ateneo de Naga, DLSU, and CIT. Therefore, the findings suggest that the individuals and the PHEIs support IT research undertakings in the Philippines. Furthermore, the results indicate that the PHEIs and their faculty members are responding to the needs of the country to build a vibrant IT research.

In terms of research topic, algorithm is the most investigated field. This is expected since IT research involves algorithms. It is interesting to note that research studies in the area of mobile computing are emerging. The term mobile was first mentioned during the 2011 conference. Since then, research projects implemented in a mobile platform have been conducted. This implies that institutions are responding to research trends.

The words "design", "systems", "management", "performance", "measurement", and "experimentation" are the most prevalent CCS concepts investigated. These all signify the investigation of the software itself. "Data" and "information" are not CCS concepts. These are author-supplied keywords. These keywords indicate that the focus of the research is not the software itself but on the data and the information it generated. Inaccurate entries in the CCS concepts resulted to three different keywords. This is evident in the words "human" and "factors", and on the phrase "human factors". Nevertheless, the findings imply that the human aspects of computing are one of the most investigated fields of study.

The author-by-number of collaborators analysis shows that researchers tend to work in small group. They prefer to have at least one collaborator but not more than 3 authors. One of the possible reasons is that collaboration brings out the best possible ideas and perspectives from the team members. Strategic management of workload and resource allocations may also explain this finding. Nevertheless, whatever may be the reason; it is apparent that they are not working in isolation.

## 6. Conclusions, Recommendations, and Limitations

This study aims to describe the IT research productivity in the Philippines. It was shown that the NCITE and the CSP were able to attract research papers from different institutions. It is found that there private HEIs dominated the list of research productive institutions. The most prolific writer is from the state-run university of the Philippines. Authors do not work in isolation but they prefer to work in a small group (2-3 members). Studies concerning algorithms are the most investigated topic and systems in mobile platforms are emerging. Thus, it can be concluded that both organizations are continuously committed to achieving their goals. Both organizations may conduct a program that could inform their members about the practices of each institution and of the individual researchers in achieving research productivity.

However, this study has its inherent limitations. They study identified the institutions, and personalities that are doing well in research. The study did not investigate the causes that make them productive in research. It is recommended that institutional practices, culture, and support, and individual motivations be investigated to shed light on this matter.

The current study did not considered international publications. Hence, future studies may investigate the research productivity of Filipino researchers in terms of global publications.

## ACKNOWLEDGMENTS

The authors are indebted to Dr. Ma. Regina Estuar, Dr. Marlene M. De Leon and John Noel Victorino. This paper is funded by the University of the East.

## REFERENCES

[1] J. L. Allen, K. Y. Huggins-Hoyt, M. J. Holosko, and H. E. Briggs (2018). African american social work faculty: Overcoming existing barriers and achieving research productivity. Research on Social Work Practice, 28(3), 309-319.

[2] J. H. Barnhill (2017). Social network analysis (SNA). Salem Press Encyclopedia. http://rizal.lib.admu.edu.ph:2048/login?url=http://search.ebscohost.com/login.aspx?direct=true&db=ers&AN=90558462&scope=site

[3] Computing Society of the Philippines (CSP). https://csp.org.ph/

[4] A. Kuzhabekova and A. Ruby (2018). Raising research productivyt in a post-Soviet higher education system: A case from Central Asia. European Education, 50, 266-282.

[5] E. P. S. Lase and Y. Hartijasti (2018). The effect of individual and leadership characteristics toward research productivity with institutional characteristics as a mediator variable: Analysis of academic lecturers in the faculty of economics and faculty of languages and arts at Univeristy of X. The South East Asian Journal Management, 12(1), 20-42.

[6] W. Lou, H. Wang, and S. Yang (2018). Chinese scholars in China and overseas: Comparative analysis on research productivity and impact. Current Science, 115(1), 49-55.

[7] F. M. Nasser-Abu Alhija and A. Majdob (2017). Predictors of Teacher Educators' Research Productivity. Australian Journal of Teacher Education, 42(11), Article 3.

[8] E. A. Obuku, J. N. Lavis, A. Kinengyere, R. Ssenono, M. Ocan, D. K. Mafigiri, F. Ssengooba, C. Karamagi and N. K. Sewankambo (2018). A systematic review on academic research productivity of postgraduate students in low- and middle-income countries. Health Research Policy and Systems, 16(86), 1-8.

[9] Philippine Society of IT Educators (PSITE). http://psite-ncr.org/index.php?option=com_content&view=article&id=1&Itemid=2

[10] R. C. Rodrigueza and M. R. J. E. Estuar (2018). Social Network Analysis of a Disaster Behavior Network: An Agent-Based Modeling Approach. In 2018 IEEE/ACM International Conference on Advances in Social Networks Analysis and Mining (ASONAM) (pp. 1100-1107). IEEE.

[11] B. Sun, M. Wang, and W. Guo (2018). The influence of grouping/non-grouping strategies upon student interaction in online forum: A social network analysis. In 2018 International Symposium on Educational Technology (ISET)(pp. 173-177). IEEE.

[12] Q.-H. Vuong, N. K. Napier, T. M. Ho, V. H. Nguyen, T.-T. Vuong, H. H. Pham, and H. K. T. Nguyen (2018). Effects Effects of work environment and collaboration on research productivity in Vietnamese social sciences: evidence from 2008 to 2017 scopus data. Studies in Higher Education, online first.